\documentclass[aps,prl,reprint,twocolumn,oneside,floatfix,sort&compress,superscriptaddress,showpacs,nobalancelastpage,longbibliography,preprintnumbers]{revtex4-1}
\usepackage{graphicx,xcolor}
\usepackage{amssymb,bbm}
\usepackage[reqno]{amsmath}

\usepackage[colorlinks,bookmarks=false,citecolor=blue,linkcolor=red,urlcolor=blue,hyperfootnotes=true]{hyperref}

\begin{document}

\title{Mott Quantum Criticality in the Anisotropic 2D Hubbard Model }

\author{Benjamin Lenz}
 \email{benjamin.lenz@theorie.physik.uni-goettingen.de}
\affiliation{Institute  for  Theoretical  Physics,  University  of  G\"ottingen, Friedrich-Hund-Platz  1,  D-37077  G\"ottingen,  Germany}
\author{Salvatore R. Manmana}
\affiliation{Institute  for  Theoretical  Physics,  University  of  G\"ottingen, Friedrich-Hund-Platz  1,  D-37077  G\"ottingen,  Germany}
\author{Thomas Pruschke}\altaffiliation{Deceased}
\affiliation{Institute  for  Theoretical  Physics,  University  of  G\"ottingen, Friedrich-Hund-Platz  1,  D-37077  G\"ottingen,  Germany}
\author{Fakher F. Assaad}
\affiliation{Institute for Theoretical Physics and Astrophysics, University of W\"urzburg, Am Hubland, D-97074 W\"urzburg, Germany}
\author{Marcin Raczkowski}
\email{marcin.raczkowski@physik.uni-wuerzburg.de}
\affiliation{Institute for Theoretical Physics and Astrophysics, University of W\"urzburg, Am Hubland, D-97074 W\"urzburg, Germany}
\affiliation{Department of Physics and Arnold Sommerfeld Center for Theoretical Physics,
             Ludwig-Maximilians-Universit\"at M\"unchen, D-80333 M\"unchen, Germany}

\date{\today}                 
\preprint{NSF-KITP-15-161}

\begin{abstract}
We present evidence for Mott quantum criticality in an anisotropic two-dimensional system of coupled Hubbard chains at half-filling.
In this scenario emerging from variational cluster approximation  and cluster dynamical mean-field theory,  
the interchain hopping $t_{\perp}$ acts as a control parameter driving the second-order critical end point $T_c$ of the 
metal-insulator transition down to zero at $t_{\perp}^{c}/t\simeq 0.2$.  
Below $t_{\perp}^{c}$, the volume of the hole and electron Fermi pockets of a compensated metal vanishes continuously at 
the Mott transition. Above $t_{\perp}^{c}$, the volume reduction of the pockets is cut off by a first-order transition.
We discuss the relevance of our findings to a putative quantum 
critical point in layered organic conductors, whose location remains elusive so far.

\end{abstract}

\maketitle


A subject of strong current interest in condensed matter physics is the metal-insulator transition (MIT)~\cite{IFT98} 
with a \emph{low} critical end point $T_c$ at which the Mott transition ceases to be first order~\cite{FMT+15,AKW+15,Hartmann15}. 
The nature of this critical end point and its universality class is a long-standing issue.  
From general considerations, one expects it to belong to the Ising universality class~\cite{Castellani79,Kotliar00}, 
similar to the liquid-gas transition, with the double occupancy playing the role of a scalar order parameter of the transition.  
A canonical example is three-dimensional (3D) Cr-doped V$_2$O$_3$ where critical exponents extracted from electrical conductivity measurements 
very close to the critical end point $T_c\simeq 450$K are consistent with the universality class of the 3D Ising model~\cite{Limelette03}. 
In contrast, similar experiments on layered $\kappa$-type charge-transfer  salts
with significantly lower $T_c\simeq 40$K have indicated unconventional Mott criticality~\cite{KMK05}. 
They stimulated subsequent experimental studies either objecting the existence of unconventional behavior~\cite{dSBS+07} or supporting  it~\cite{KMK09}.
Theoretical scenarios of the two-dimensional (2D) Mott transition are also controversial, ranging from ordinary Ising universality~\cite{PFF+08, BdSL10,ST12} 
to unconventional critical exponents~\cite{SWGK11}.

Recently, the question of the nature of the 2D MIT transition has been raised again as new experiments on $\kappa$-type 
and palladium dithiolene organic conductors support either unconventional criticality~\cite{FMT+15} or 2D Ising criticality~\cite{AKW+15}, respectively. 
As the conductors studied in Ref.~\onlinecite{FMT+15} possess low-$T$ ground states with various broken symmetries,
the unconventional Mott criticality seems to be generic and unrelated to the proximity to symmetry broken states. 
Instead, the fact that the critical end point $T_c$ is relatively low  suggests quantum effects to become important and necessitates a physical picture 
contrasting the one building on classical phase transitions~\cite{Imada05,Imada07,IMY10}.  
Furthermore, a possible support for the 2D Mott quantum criticality comes from the dynamical mean-field theory (DMFT)~\cite{Terletska11,Vucicevic15},
which reveals  unexpected scaling behavior of the resistivity curves in the high-$T$ crossover region $T\gg T_c$. 
A stringent test of the link between this scaling behavior and the quantum criticality appears, however, to be impossible since the latter is masked in the half-filled 
2D Hubbard model by the low-$T$ coexistence dome~\cite{Imada03,PHK08,BKS+09,Gull09}. 
Moreover, various numerical studies find that $T_c$ remains finite also in the presence of lattice 
frustration~\cite{Parcollet04,Kyung06,WYTI06,Kawakami08,Koga09,Liebsch09,Wessel15}.  
Finally, while the effective suppression of $T_c$ can be achieved with disorder, it requires the proper treatment of Anderson localization 
effects~\cite{Byczuk94,Aguiar05,Braganca15}.

In this Letter, we propose a different route to account for a low critical end point $T_c$ of the MIT. 
Considering the fact that quantum fluctuations are enhanced in low-dimensional systems with spatial anisotropy, 
we investigate, using two complementary state-of-the-art numerical techniques, the effect of anisotropic hopping amplitudes 
and try to locate the putative quantum critical point at $T=0$ in the phase diagram. 
 
\paragraph{Model and methods.} 
We study the frustrated Hubbard model on a square lattice with an anisotropic hopping at half-filling:
\begin{equation}
\mathcal{H}=-\sum_{\pmb{ij},\sigma}t^{}_{\pmb{ij}}
   c^{\dag}_{{\pmb i}\sigma}c^{}_{{\pmb j}\sigma} +
   U\sum_{\pmb i}n^{}_{{\pmb i}\uparrow}n^{}_{{\pmb i}\downarrow} 
   -\mu\sum_{\pmb i,\sigma}n_{{\pmb i}\sigma},
\label{eq:Hubb}
\end{equation}
with chemical potential $\mu$ and Coulomb repulsion $U$. The hopping $t_{\pmb{ij}}$ is $t$ along the chains and $t_\perp$ between the chains. 
By tuning the ratio $t_\perp/t$ from 0 to 1, we bridge the limit of uncoupled one-dimensional (1D) Hubbard chains ($t_{\perp}=0$) 
and the isotropic 2D lattice ($t_{\perp}/t=1$). 
In order to remove the perfect nesting antiferromagnetic (AFM) instability of the Fermi surface (FS)~\cite{Hofstetter98} also in the 2D regime,
which would lead to an insulating state at any finite value of $U$~\cite{NSST08, Held15}, we add geometrical frustration via next-nearest-neighbor hopping 
$t' = -t_\perp/4$.

\begin{figure}[t!]
\centering
 \includegraphics[width=0.99\columnwidth]{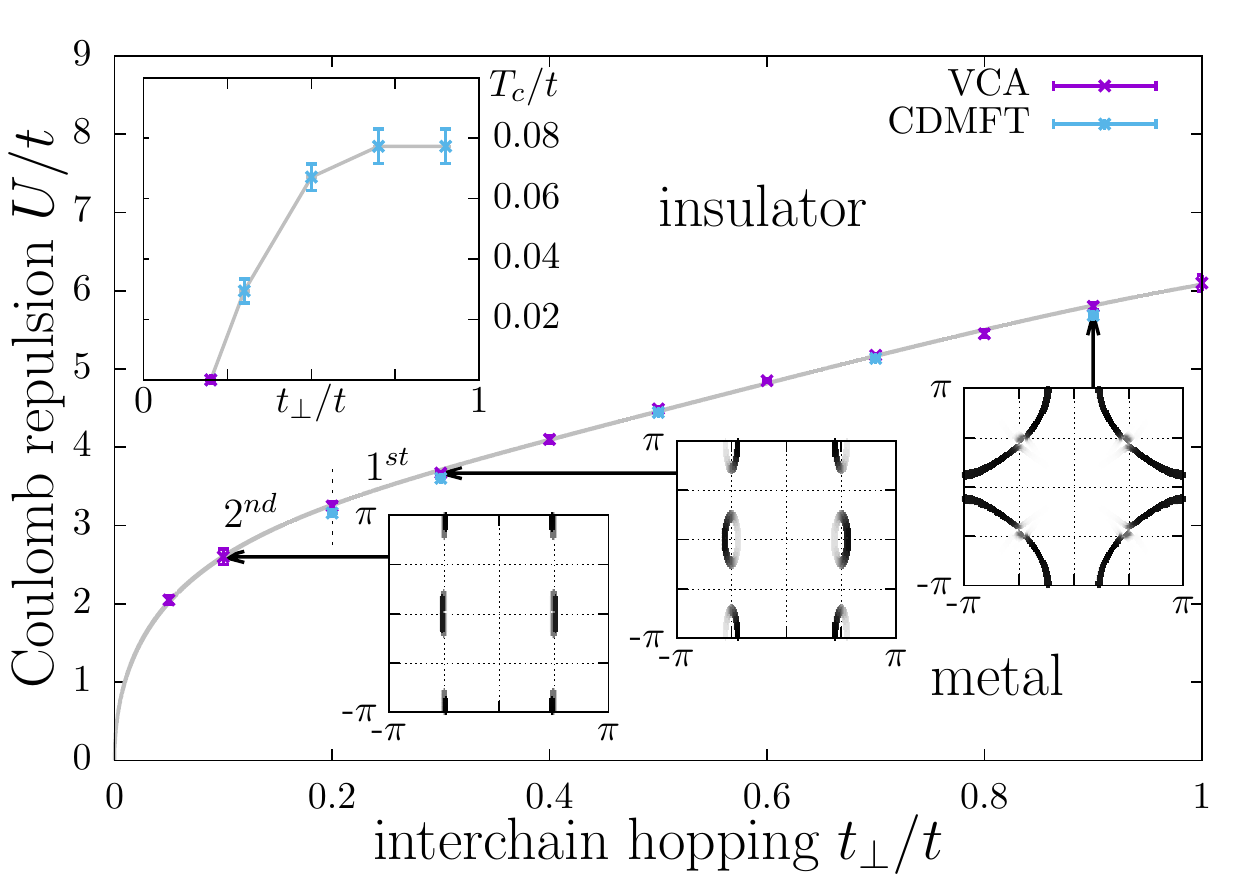} 
 \caption{Metal-insulator phase diagram of the half-filled Hubbard model (\ref{eq:Hubb}) as obtained by the zero-temperature 
VCA and CDMFT at $T=t/40$.   
(Top inset) The combined VCA and CDMFT estimate for the critical temperature $T_c$ terminating the first-order MIT; 
$T_c$ is driven down to zero at $t_{\perp}^c/t\simeq 0.2$, thus providing evidence for the quantum critical nature of the MIT.
(Bottom insets) FS topology close to the critical interaction $U_c$ in different regions of the phase diagram indicated by arrows.
}
 \label{fig:PD}
\end{figure}

The results are obtained by two complementary quantum cluster techniques~\cite{Maier05}, which can both be described within the framework of self-energy 
functional theory~\cite{Pot03a}. In the cluster extension of DMFT (CDMFT)~\cite{KSPB01}, a cluster of $N_c$ interacting impurities is dynamically 
coupled to an effective bath. The impurity problem is solved using the quantum Monte Carlo (QMC) Hirsch-Fye solver and coupling to the bath is 
determined self-consistently. 
To also make the study computationally manageable down to the lowest temperature $T=t/40$ in the 2D regime, where the sign problem hampers the 
usage of the QMC solver on larger clusters, we use a $2\times2 $ plaquette cluster. 
The $2\times2 $ cluster is a minimal unit cell which allows one to capture both the 1D umklapp scattering process opening a gap in the  
half-filled band~\cite{Capone04} and short-range 2D AFM spin fluctuations. 
To trace the Mott transition at zero temperature, we use the variational cluster approximation (VCA)~\cite{PAD03,Pot03} with a $2\times2 $ cluster and one additional 
bath site per correlated site as a reference system~\cite{BKS+09}, i.e., an effective eight-site cluster. 
In VCA, the grand potential $\Omega$ is approximated by the self-energy functional (SEF) at its saddle point. 
As variational parameters, we choose the hybridization $V$ between correlated and bath sites and chemical potentials of the reference 
system $\mu^{\prime}$ and the lattice system $\mu$, respectively~\footnote{Note that variation of the hopping terms on the 
cluster $t^{\prime}$ and $t_{\perp}^{\prime}$ did not lead to qualitative differences compared to a variation of $\mu, \mu^{\prime}$ and $V$ only. 
They were therefore chosen to be equal to their lattice analogues and not treated as additional variational parameters.}.

\begin{figure}[t!]
\centering
 \includegraphics[width=0.99\columnwidth]{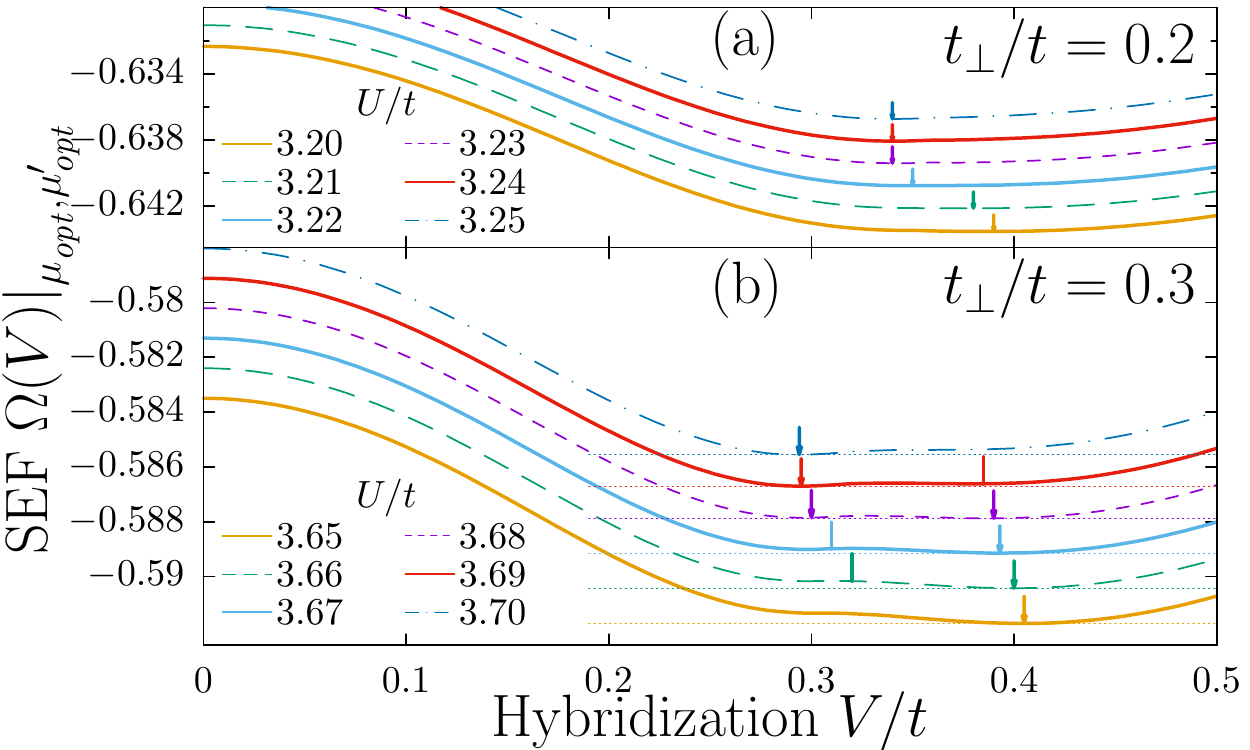} 
 \caption{VCA self-energy functional $\Omega$  in the proximity of MIT as a function of hybridization $V/t$ for 
 (a)  $t_{\perp}/t=0.2$ and (b) $t_{\perp}/t=0.3$. 
  In (b) stable minima are indicated by thick arrows; thinner ones mark unstable solutions.}
  \label{fig:SEF}
\end{figure}

\paragraph{Phase diagram.} Our main results are summarized in the ground-state phase diagram in Fig.~\ref{fig:PD}. 
It shows our estimate of the critical interaction strength $U_c$ at which the system undergoes a transition between Mott insulating and metallic phases 
in the full range between the 1D and 2D regimes. 
In agreement with the exact  Bethe ansatz solution~\cite{LiebWu68} and bosonization approach~\cite{Giamarchi_book},  VCA yields the Mott phase 
for any $U>0$ in the 1D limit~\cite{BHP08}. As shown in Fig.~\ref{fig:PD}, this changes dramatically upon coupling the chains: 
single-particle hopping  $t_{\perp}$  shifts the critical interaction $U_c$ towards a finite value, thus enabling the interaction-driven MIT. 
Initially, $U_c$ increases steeply with $t_{\perp}$ and then continues to grow nearly linearly, as expected for the MIT controlled by the ratio of 
Coulomb interaction to kinetic energy gain. 
For $t_{\perp}/t> 0.2$, the MIT line is found to be first-order consistent with former studies of the frustrated 2D Hubbard 
model~\cite{Parcollet04,Kyung06,WYTI06,Kawakami08,Koga09,Liebsch09,Wessel15}.
In contrast, in the strongly anisotropic case with $t_{\perp}/t\leq 0.2$, it marks a smooth metal-insulator crossover down to $T=0$. 
This is supported by a systematic reduction of the critical end point $T_c$ identified within 
CDMFT (see the inset of Fig.~\ref{fig:PD}). All of these aspects consistently suggest that $t_{\perp}$ is a control parameter which tunes the nature of the Mott transition 
from strong first order in the 2D limit to continuous at $t_{\perp}^{c}/t\simeq 0.2$.
We complement the phase diagram by showing the change of the FS topology when tuning $t_\perp$ for values of the interaction close to $U_c$.
Two main features come into play: (i) finite $t_\perp$ leads to a warping of the 1D FS and, in the presence of interactions, to the formation of hole and 
electron Fermi pockets of a higher-dimensional compensated metal~\cite{Essler05}, 
and (ii) for values $t_\perp/t\gtrsim 0.7$  the compensated metal structure of the FS disappears going over 
to a conventional large FS which coincides with the topological Lifshitz transition of the noninteracting FS.

\paragraph{Obtaining the phase diagram.} We now describe the numerical results which lead us to the above phase diagram. VCA provides the possibility 
of identifying and tracing competing phases by analyzing the self-energy functional  $\Omega(\mu,\mu^{\prime},V)$. 
For the interchain coupling $t_{\perp}/t=0.2$, we cannot resolve two disjoined SEF minima and the value of $V$ is thus expected to change 
continuously across the critical interaction $U_c$; see Fig.~\ref{fig:SEF}(a).  In contrast, for $t_{\perp}/t\gtrsim0.3$, the SEF has four 
saddle points, of which two correspond to stable phases close to the phase transition: one corresponding to the metallic, 
the other to the insulating solution. These stationary points of $\Omega(\mu,\mu^{\prime},V)$ are maxima with respect to $\mu$ and $\mu^{\prime}$, 
but minima with respect to hybridization strength $V$. The existence of two distinct minima in the SEF landscape shown in Fig.~\ref{fig:SEF}(b) 
results in a jump of hybridization $V$ when tuning across $U_c$, and thus signals the first-order nature of the MIT.

\begin{figure}[t!]
\centering
\includegraphics[width=0.99\columnwidth]{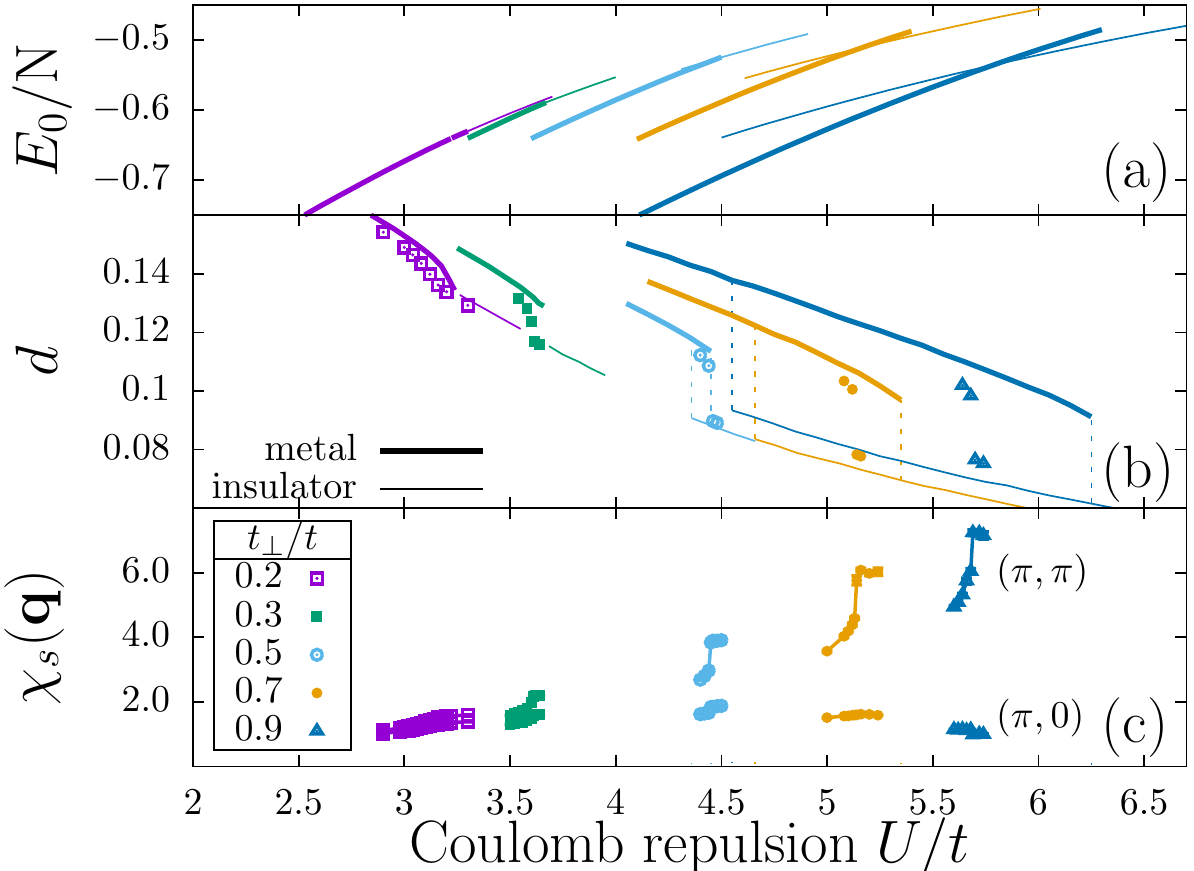}
 \caption{(a) VCA ground-state energy $E_0$ as a function of Coulomb repulsion $U/t$.
   Filled (dashed) lines indicate metallic (insulating) solutions for  
   $t_{\perp}/t=0.2,0.3,0.5,0.7$, and 0.9 (from left to right).
 (b) VCA double occupancy $d$ across the MIT at $T=0$; symbols stand for CDMFT results at $T=t/40$. 
 (c) Cluster spin susceptibility $\chi_s({\pmb q})$ within 
  CDMFT at $T=t/40$; the upper curves correspond to spin fluctuations  at the AFM wave vector ${\pmb q}=(\pi,\pi)$ and 
 the lower ones to remnant 1D fluctuations at ${\pmb q}=(\pi,0)$.}
 \label{fig:double_occup}
\end{figure}

Next, we focus on the ground-state energy $E_0$ and the double occupancy $d$. The latter is obtained as the derivative of the grand potential $\Omega$ 
with respect to Coulomb repulsion $d=\frac{\partial\Omega}{\partial U}$.
In the quasi-2D region we identify a clear kink in the ground-state energy $E_0$; see Fig.~\ref{fig:double_occup}(a). It arises from a level crossing 
of the insulating and metallic solutions and gives rise to a jump in the double occupancy $d$  at $U_c$, as shown in Fig.~\ref{fig:double_occup}(b). The latter 
exhibits hysteresis in the region with two solutions, as expected for the first-order transition. Although a weak kink in $E_0$ is also resolved 
for intermediate values of $t_{\perp}$, both the coexistence region and the jump in the double occupancy shrink and vanish at  $t_{\perp}^{c}/t\simeq 0.2$~\cite{SupplMat}. 
The absence of a jump in $d$ and a single minimum in the SEF yield strong evidence for the continuous nature of the MIT. 
A similar scenario emerges within a finite-temperature CDMFT: while a clear jump in 
$d=\tfrac{1}{N_c}\sum_{\pmb{i}}\langle n^{}_{{\pmb i}\uparrow}n^{}_{{\pmb i}\downarrow}\rangle$ 
is found in the quasi-2D regime, it  gradually decreases when reducing $t_{\perp}$ and finally converts into a crossover at $t_{\perp}^{c}/t=0.2$. 
It remains smooth down to our lowest temperature $T=t/40$; see Fig.~\ref{fig:double_occup}(b).
As shown in Fig.~\ref{fig:double_occup}(c), the level crossing in the ground state is also reflected in the spin sector and produces a jump 
in the cluster spin susceptibility 
$\chi_s({\pmb q})=
\tfrac{1}{N_c}\int_{0}^{\beta}d\tau\sum_{\pmb{ij}} e^{i{\pmb q}\cdot({\pmb i}-{\pmb j})}
\langle\pmb{ S_i}(\tau)\pmb{S_j}(0)\rangle$
at the AFM wave vector ${\pmb q}=(\pi,\pi)$. 
In contrast, no distinction between the response in $\chi_s({\pmb q})$ at ${\pmb q}=(\pi,0)$ and ${\pmb q}=(\pi,\pi)$  wave vectors   
at $t_{\perp}/t=0.2$  indicates that remnant 1D effects play an important role in the weakly coupled regime.  

\begin{figure}[t!]
\centering
 \includegraphics[width=0.49\columnwidth]{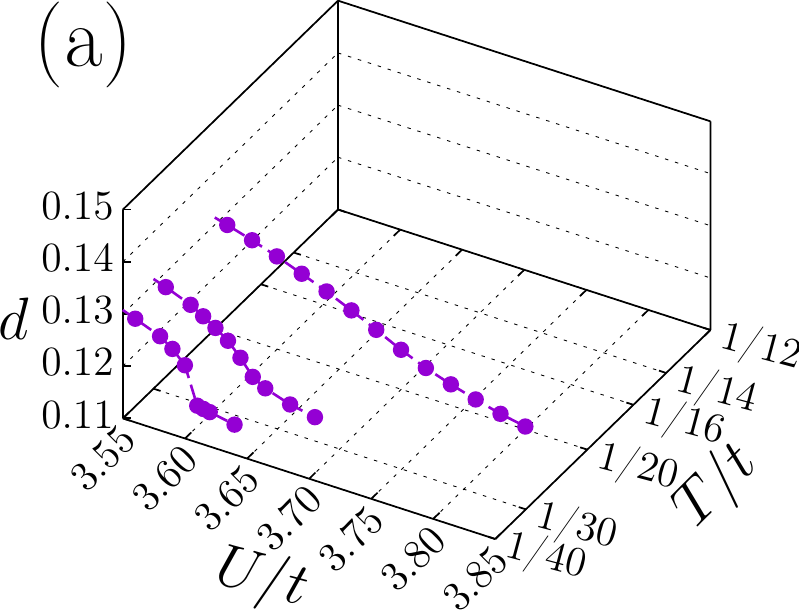} 
 \includegraphics[width=0.49\columnwidth]{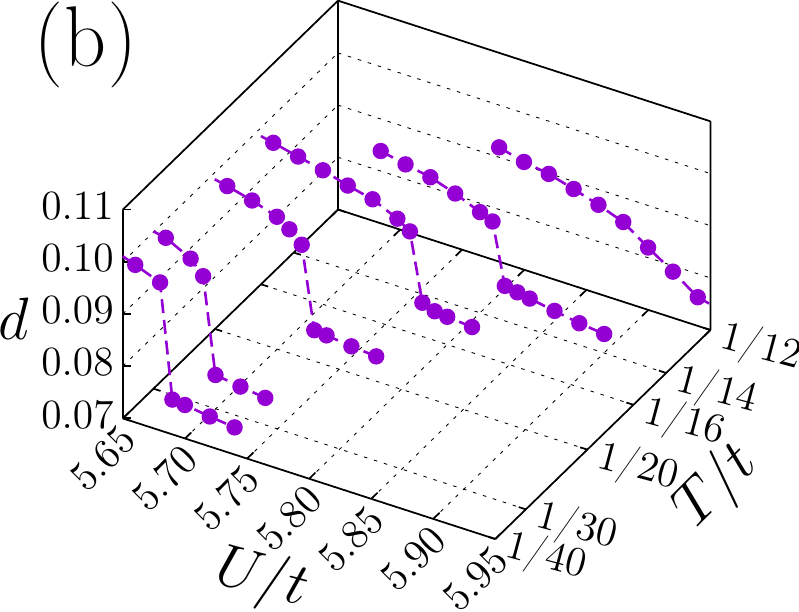} 
 \caption{Double occupancy $d$  as a function of interaction $U/t$ obtained in CDMFT for 
 (a) $t_{\perp}/t=0.3$ and (b) $t_{\perp}/t=0.9$. The low-$T$ jump in $d$ signaling the first-order MIT   
  turns into a crossover above the critical end point $T_c$.}
 \label{fig:D_T}
\end{figure}

We turn now to finite-temperature consequences of the continuous MIT seen at $T=0$. The estimate of $T_c$ at a given $t_{\perp}$  was obtained 
by monitoring $d$ as a function of $U/t$ at fixed $T$; see Fig.~\ref{fig:D_T}. The low-$T$ jump in $d$ signaling the first-order MIT remains up to 
$T_c$ and turns into a smooth crossover above $T_c$.
As shown in Fig.~\ref{fig:D_T}(a), for small $t_{\perp}/t=0.3$ a smooth behavior in $d$ is already recovered at $T=t/30$. 
In contrast, for $t_{\perp}/t=0.9$, the jump converts into a crossover at much higher temperature $T=t/12$. By repeating the above analysis 
for intermediate values of $t_{\perp}$, we extracted the $t_{\perp}$ dependence of the critical temperature $T_c$ 
(see the inset in Fig.~\ref{fig:PD}) consistent with a continuous reduction of the jump in the double occupancy~\cite{SupplMat}.

\begin{figure}[t!]
\centering
 \includegraphics[width=0.99\columnwidth]{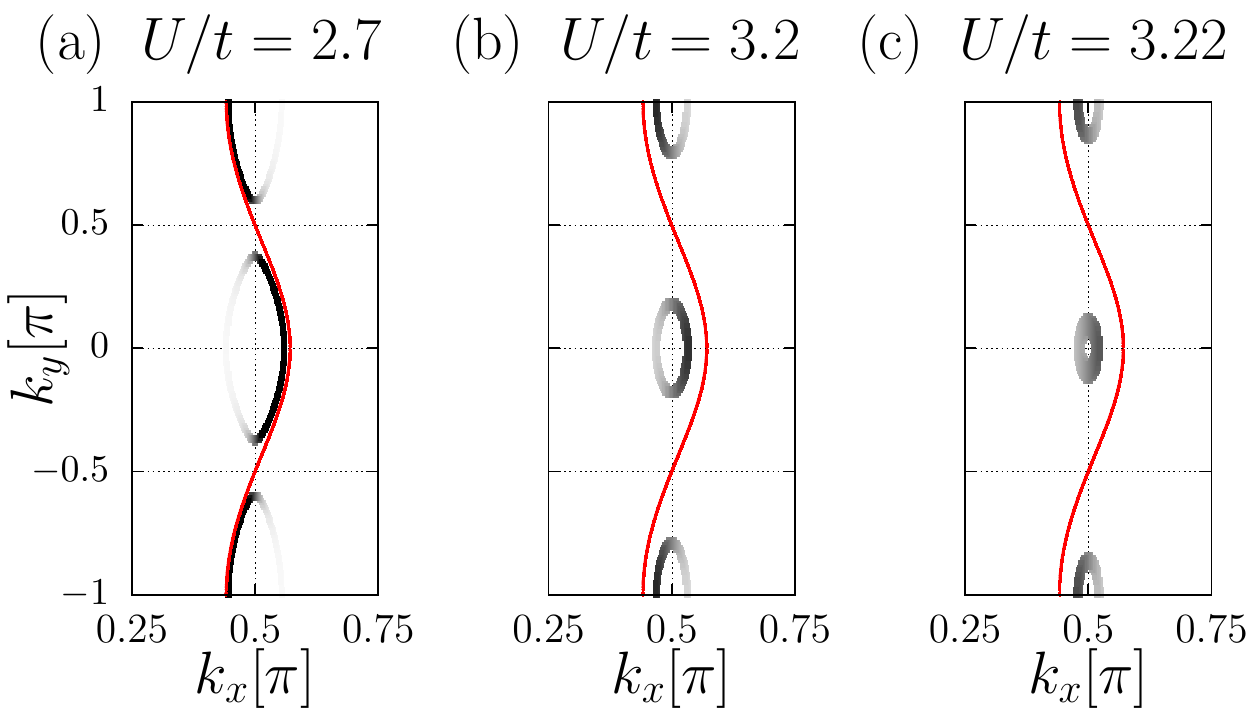} 
 \caption{Evolution of the FS with electron and hole pockets (see the text) for $t_{\perp}/t=0.2$ when approaching $U_c/t\simeq 3.22$ from below: 
 (a) $U/t=2.7$, (b) $U/t=3.2$, and (c) $U/t=3.22$. Red solid lines show the noninteracting dispersion.}
 \label{fig:Pock}
\end{figure}

\begin{figure}[b!]
\centering
 \includegraphics[width=0.99\columnwidth]{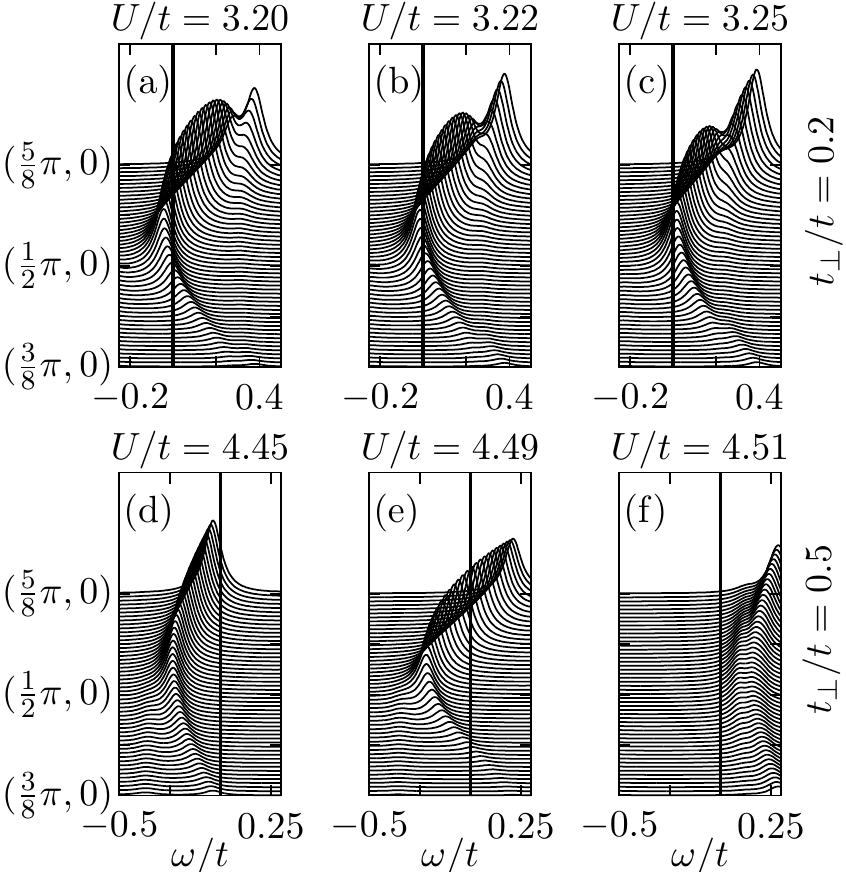} 
\caption{Low-energy part of the single-particle spectral function  $A({\pmb k},\omega + i\eta)$ for 
$t_{\perp}/t=0.2$ (top panels)  and $t_{\perp}/t=0.5$ (bottom panels) obtained within VCA at $T=0$, $\eta=0.05$. 
The FS pockets are found (a),(d) for $U<U_c$ and (b),(e) at $U\lesssim U_c$; 
in the insulator at (c),(f) $U>U_c$, their disappearance signifies the vanishing of the FS, and hence the MIT. 
}
\label{fig:Akw}
\end{figure}

\paragraph{Spectral function.} 
To elucidate the microscopic origin of the continuous Mott transition for $t_{\perp}/t\lesssim 0.2$, we calculate the single-particle spectral 
function $A({\pmb k},\omega)=-\frac{1}{\pi}\text{Im} G({\pmb k},\omega)$,  where $G({\pmb k},\omega)$ is the lattice Green's function. 
Figure~\ref{fig:Pock} shows the evolution of $A({\pmb k},\omega)$ upon increasing the interaction $U$ at fixed $t_{\perp}/t=0.2$. 
In agreement with random-phase approximation studies~\cite{Essler02,Penc11}, we find that the destruction of the FS starts at  
momenta ${\pmb k} = (\pi/2,\pm\pi/2)$, where the interchain hopping matrix elements vanish. As a result, a compensated metal structure of 
the FS is formed with elliptic electron and hole pockets around the ${\pmb k}=(\pi/2,0)$ and $(\pi/2,\pi)$ points.
A striking feature of the pockets is their symmetric form contrasted with pockets found in coupled  \emph{spinless} 
fermionic chains~\cite{Berthod06}. We ascribe this symmetry to quasiparticle scattering off short-range 
1D spin fluctuations with ${\pmb q}=(\pi,0)$.  On one hand, at intermediate interaction strengths, the main part of the FS carrying most 
of the quasiparticle weight follows closely the noninteracting FS. 
On the other hand, the pockets shrink in size and become very shallow close to $U_c$; see Figs.~\ref{fig:Akw}(a)-~\ref{fig:Akw}(c).
The continuous vanishing of the volume of Fermi pockets at $U_c$ implies the second-order nature of the MIT.
Since the inverse width of the hole or electron pockets defines a characteristic length scale, $\xi$, one should be able to extract the correlation 
length exponent, $\nu$, from a careful study of the critical behavior of the volume of the pocket as one approaches $U_c$~\cite{IFT98,SupplMat}.
In contrast, the volume reduction of the pockets is cut off by a first-order transition for larger $t_{\perp}$; 
cf. Figs.~\ref{fig:Akw}(d)-~\ref{fig:Akw}(f) with $t_{\perp}/t=0.5$.

\paragraph{Discussion and outlook.} 
Let us relate our findings to recent experiments on the organic conductors with a half-filled band~\cite{FMT+15}.
Both $\kappa$-(BEDT-TTF)$_2$Cu$_2$(CN)$_3$ and EtMe$_3$Sb[Pd(dmit)$_2$]$_2$ (BEDT-TTF=bis(ethylenedithio)-tetrathiafulvalene, 
dmit=1,3-dithiole-2-thione-4,5-dithiolate, Me=CH$_3$, Et=C$_2$H$_5$) are thought to be layered systems 
with H\"uckel parameters close to an equilateral triangular lattice~\cite{Kanoda_rev}. Instead, careful \emph{ab initio} calculations 
for the latter show an appreciable 1D anisotropy with the ratio of interchain to intrachain transfer around 0.82~\cite{Imada12}. 

We took this asymmetry into consideration: using VCA at $T=0$ and CDMFT at finite $T$ 
we have found strong evidence for Mott quantum criticality in coupled Hubbard chains at half-filling.
In this scenario, the interchain hopping $t_{\perp}$ acts as control parameter driving the second-order critical end point $T_c$ 
of the interaction-driven MIT down to zero in the presence of strong anisotropy.
At a threshold value of $t_{\perp}^{c}/t\simeq 0.2$, the volume of Fermi pockets shrinks to zero. The resulting MIT is \emph{continuous} 
without a detectable jump in the double occupancy or a visible coexistence region in the SEF. 
In contrast, the volume reduction of the pockets is only partial at larger $t_{\perp}$: 
the jump in the double occupancy and the existence of two distinct degenerate minima in the SEF are consistent with a first-order transition. 

The continuous MIT at $T=0$ offers a possibility for understanding the scaling behavior of resistivity curves in the high-$T$ crossover region $T\gg T_c$  
usually attributed to (hidden) 2D Mott quantum criticality ~\cite{Terletska11,Vucicevic15}.  
It is an interesting question as to whether the quantum critical behavior emerges also in coupled  \emph{spinless} fermionic chains  displaying 
similar FS breakup into Fermi pockets~\cite{Berthod06}. 

While the $2\times 2$ plaquette cluster used is known to overestimate the singlet formation~\cite{Maier05}, we expect the unveiled quantum 
critical behavior to be robust. Indeed,  former CDMFT studies on larger clusters of up to 16 sites have provided evidence for a continuous 
dimensional-crossover-driven MIT down to the lowest accessible temperatures~\cite{RA12}. 
We believe that this scenario is not restricted to quantum cluster descriptions of the system but should also emerge in lattice simulations, 
provided that the range of AFM spin fluctuations is reduced, e.g., by geometrical frustration or disorder~\cite{Byczuk09,Furukawa15}.
This leads, however, to a severe sign problem which renders lattice QMC simulations very expensive~\cite{RAP15}. 
In this respect, a promising route avoiding the main shortcomings of QMC is offered by tensor network methods~\cite{Orus14} adapted recently to fermionic 
systems~\cite{Corboz09,Corboz10}.
Our results provide a novel axis in the phase diagram along which $T_c$ can be tuned to zero. It remains to be verified if this quantum critical behavior 
can explain fingerprints of the unconventional Mott criticality observed in layered organic conductors. 

\acknowledgments
We thank M. Imada and A.M. Tsvelik for the insightful discussions. 
F.F.A. and S.R.M. would like to thank the KITP, where part of this work was carried out, for hospitality (Grant No. NSF PHY11-25915).
We acknowledge the computer support of the GWDG, the GoeGrid project, and the J\"ulich Supercomputing Centre and financial support from 
DFG Grants No. PR298/15-1 (FOR 1807) and No. AS120/8-2 (FOR 1346) as well as from FP7/ERC Starting Grant No. 306897.

\paragraph{Note added.} While this manuscript was under review, one of the co-authors, Thomas Pruschke, passed away.


\clearpage 
\setcounter{figure}{0}
\renewcommand{\thefigure}{\arabic{figure}S}

\makeatletter
     \@addtoreset{figure}{section}
\makeatother

\onecolumngrid
\begin{center}
\Large{\bf Supplemental Material for: Mott Quantum Criticality \\ in the Anisotropic 2D Hubbard Model }
\end{center}
\twocolumngrid

\subsection{Phase diagram} 

In VCA we discriminate metal against insulator by looking at the electron filling $\langle n\rangle$ as a function of chemical potential $\mu$. 
It shows a plateau at half filling for large Coulomb repulsion $U$, which corresponds to a vanishing charge compressibility 
$\kappa=-\frac{\partial\langle n\rangle}{\partial\mu}$ and indicates an insulator, whereas for a metal no plateau occurs.

\subsection{Double occupancy}

\begin{figure}[b!]
\centering
 \includegraphics[width=0.49\columnwidth]{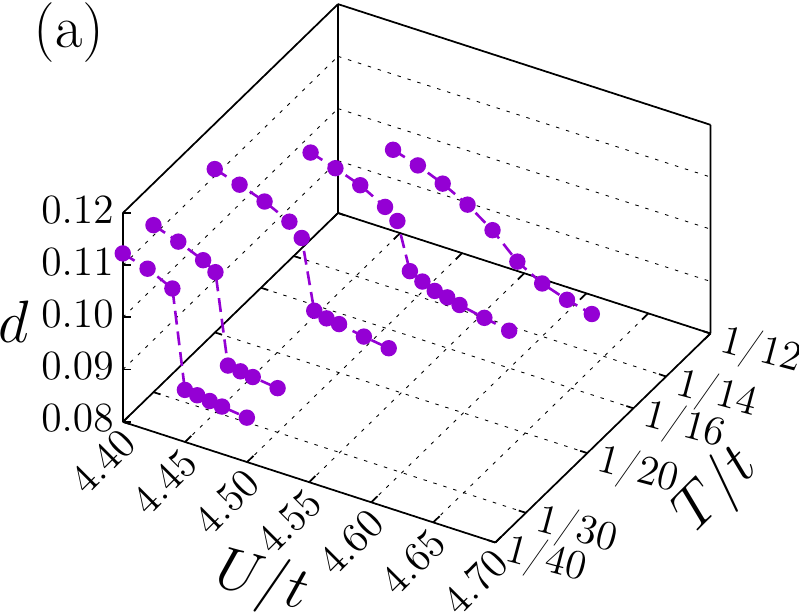} 
 \includegraphics[width=0.49\columnwidth]{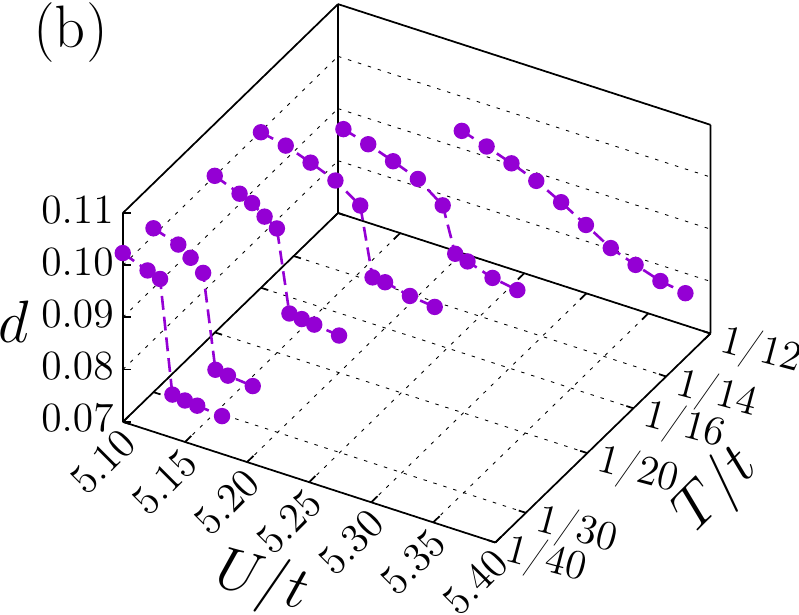} 
 \caption{(Color online) Double occupancy $d$ as a function of Coulomb repulsion $U/t$ obtained within CDMFT for: (a) $t_{\perp}/t=0.5$ and (b) $t_{\perp}/t=0.7$. 
 The low-$T$ jump in $d$ signaling the first-order MIT  turns into a crossover above the critical endpoint $T_c$.  
} 
\label{fig: ExtTc}
\end{figure}

Within CDMFT with the QMC solver the model system can be analyzed at finite-temperatures $T$. This allows one to estimate the $t_{\perp}$ dependence 
of the critical endpoint temperature $T_c$. In Fig.~\ref{fig: ExtTc} we show the double occupancy $d$ as a function of Coulomb repulsion $U/t$ for 
intermediate values of the interchain hopping $t_{\perp}/t=0.5$ and $t_{\perp}/t=0.7$ thus complementing Fig.~4 in our Letter.

\begin{figure}[b!]
\centering
 \includegraphics[width=0.9\columnwidth]{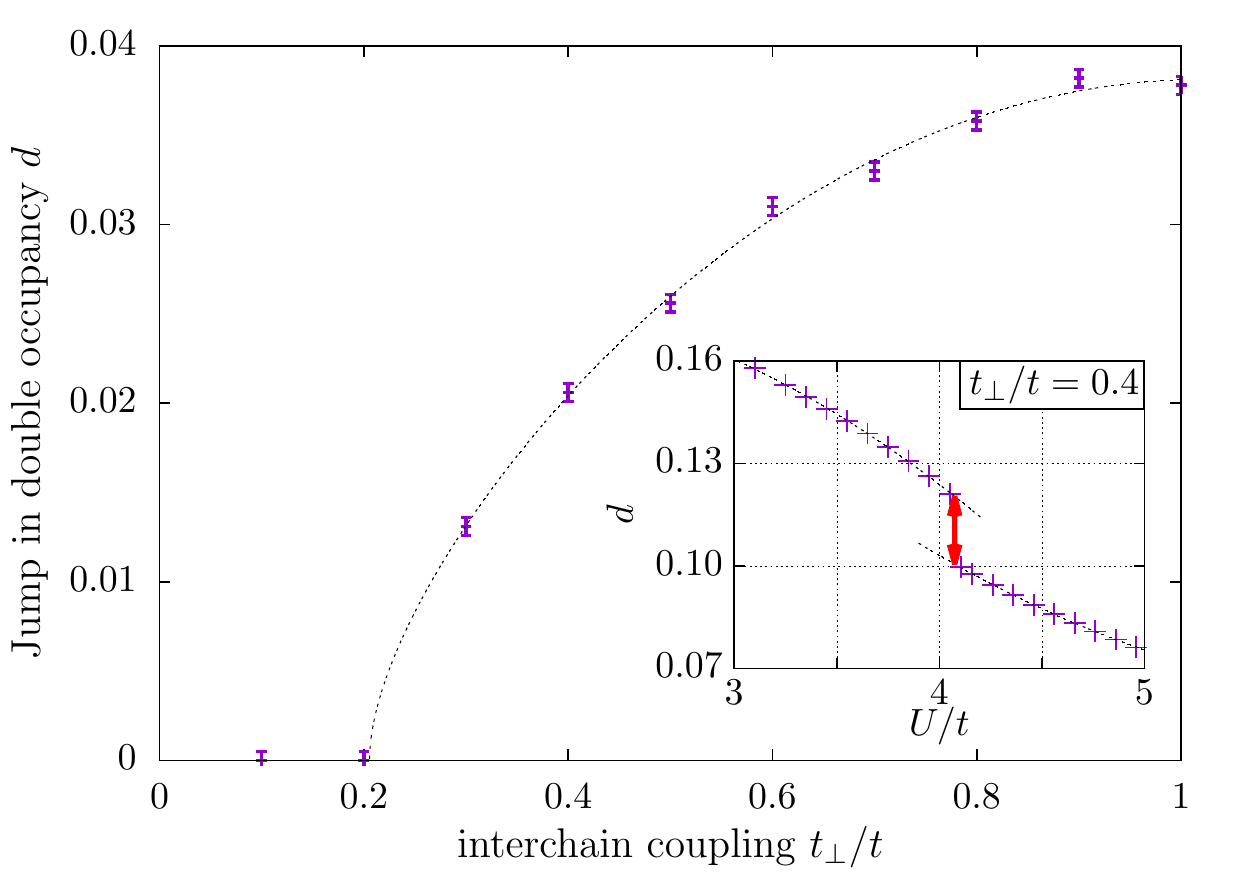} 
 \caption{(Color online) Size of the jump in double occupancy in VCA at the metal-insulator transition at $T=0$. 
 The dashed line is a rough fit and only meant as a guide to the eye. 
 For $t_{\perp}/t\lesssim0.2$ we cannot identify a jump in $d$ anymore and the transition seems to be continuous.}
 \label{fig: D_Jump}
\end{figure}

As there is no jump in $d$ within CDMFT for small interchain couplings $t_{\perp}/t\lesssim0.2$ down to the lowest accessible  temperature $T=t/40$, 
VCA is used to investigate the system at $T=0$. 
Figure \ref{fig: D_Jump} shows the size of the jump in $d$ as a function of interchain coupling. For intermediate and large $t_{\perp}$, the metallic and 
insulating solutions both exist in close vicinity of the transition and the jump can be read off immediately. In the case of small  
$t_{\perp}$, it is necessary to fit double occupancies of the metallic and insulating solutions to read off the jump at the transition 
(see inset in Fig. \ref{fig: D_Jump}). Although an exponentially small jump for $t_{\perp}/t\lesssim 0.2$ can still not be ruled out in this way, 
a fit of the jump sizes for $t_\perp/t>0.2$ agrees with a vanishing jump size for small $t_{\perp}$.

\subsection{Fermi surface pockets}

\begin{figure}[t!]
\centering
 \includegraphics[width=0.9\columnwidth]{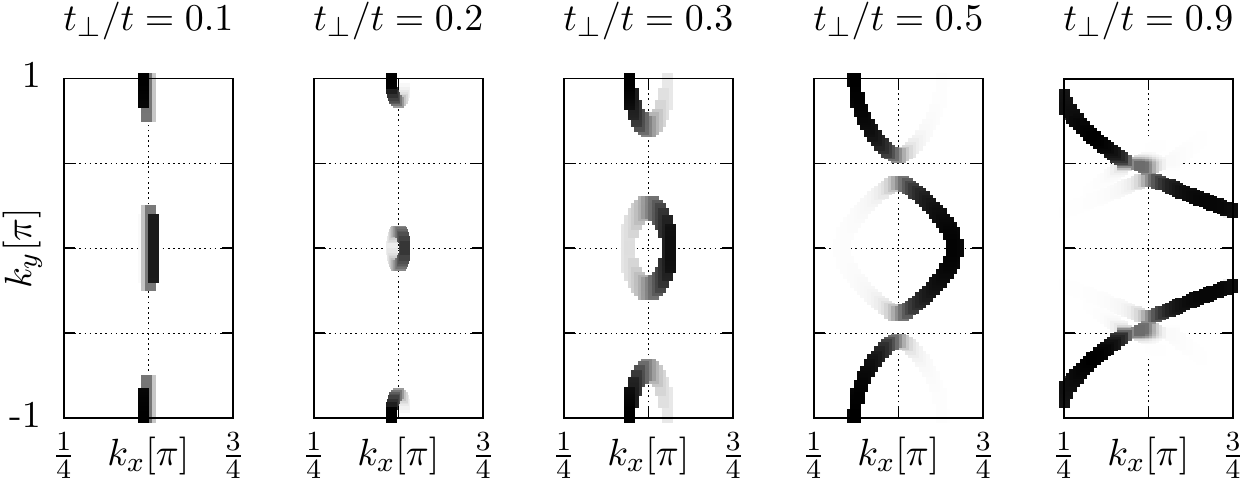}
 \caption{Fermi surface topology for several values of the interchain coupling  $t_{\perp}/t$ close to  the critical interaction strength $U_c$. 
  For $t_{\perp}/t> 0.7$ the compensated metal structure of the FS with hole and electron pockets disappears going over to a conventional large Fermi surface.}
 \label{fig: FSP}
\end{figure}

\begin{figure}[b!]
\centering
 \includegraphics[width=0.85\columnwidth]{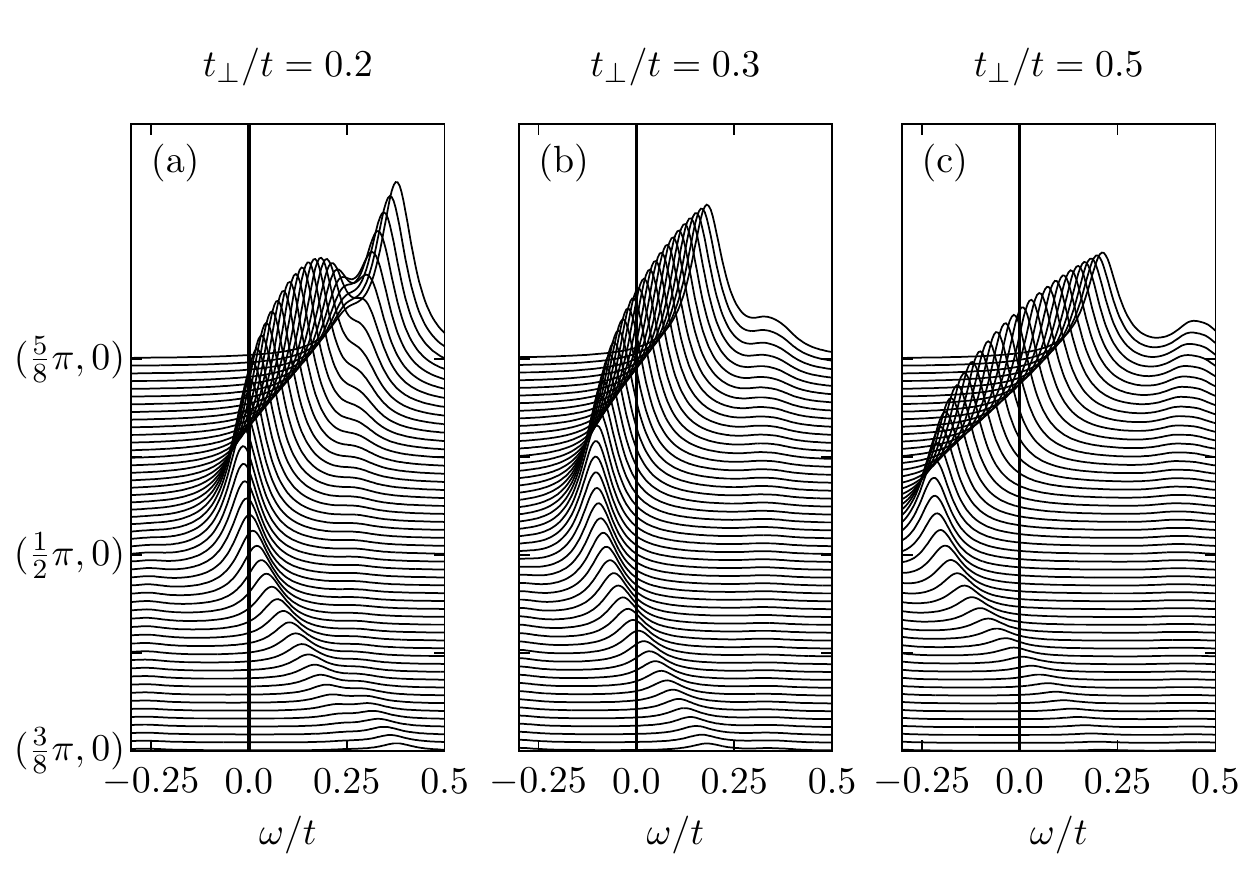}
 \includegraphics[width=0.85\columnwidth]{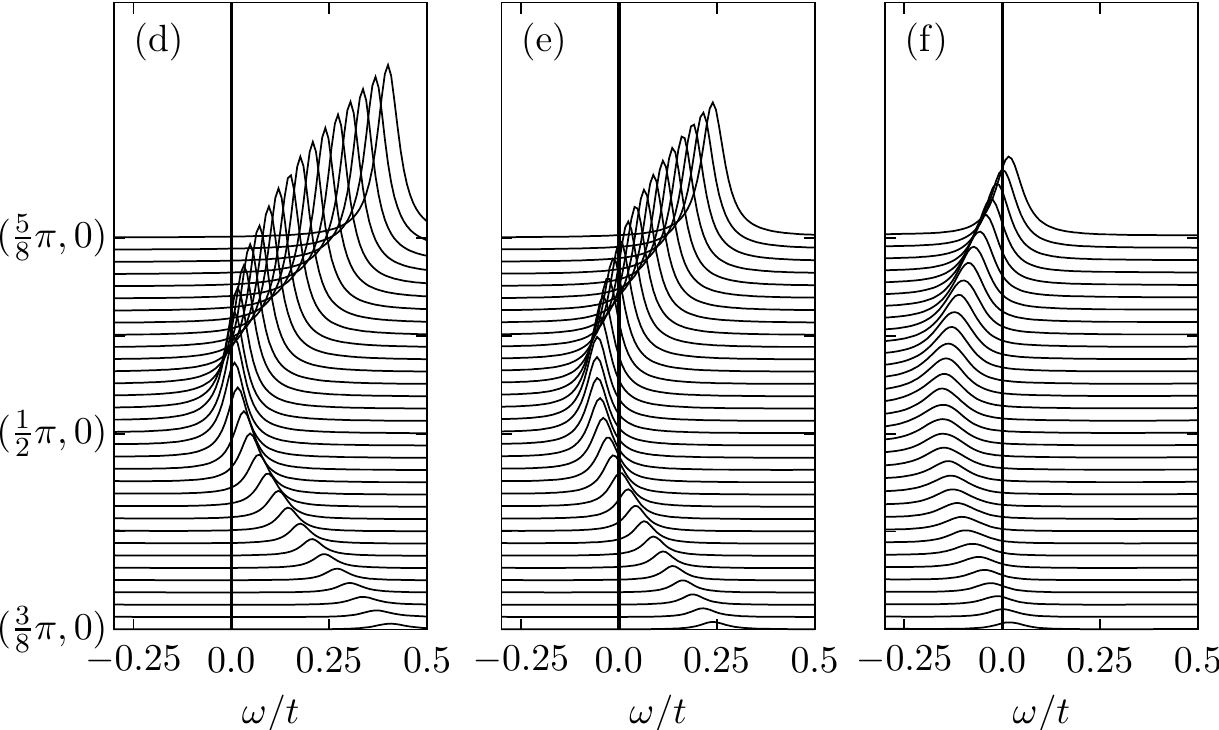}
 \caption{
Low-frequency part of the single-particle spectral function  $A({\pmb k},\omega + i\eta)$ in the metallic phase close to the critical interaction $U_c$ 
for different $t_{\perp}$ obtained within VCA at $T=0$ (top) and  CDMFT at $T=t/40$ (bottom): 
(a,d) $t_{\perp}/t=0.2$;  (b,e) $t_{\perp}/t=0.3$, and  (c,f) $t_{\perp}/t=0.5$.
In VCA small broadening $\eta=0.05$ was used; stochastic analytic continuation~\cite{Bea04S} of the imaginary-time QMC data was applied within CDMFT.
}
 \label{fig: FSPock_Cut}
\end{figure}

Figure~\ref{fig: FSP} shows the FS for different values of $t_{\perp}/t$ close to the critical value $U_c$. 
When comparing to the tight-binding limit, the main effect of the interaction is the formation of  additional shadow-like bands, and the creation of gaps 
in the originally continuous structure. Both effects lead to the formation of the compensated metal structure of the FS with hole and electron pockets
in the region $t_{\perp}/t <0.7$.
In the case of a continuous transition for $t_{\perp}/t\le 0.2$, the volume of the hole and electron Fermi pockets shrinks to zero at the critical 
interaction strength $U_c$ as found in both VCA and CDMFT, see Figs.~\ref{fig: FSPock_Cut}(a,d). For larger $t_\perp$ the pockets retain 
a finite volume up to $U_c$, see Figs.~\ref{fig: FSPock_Cut}(b,e) with $t_{\perp}/t= 0.3$  and \ref{fig: FSPock_Cut}(c,f) with $t_{\perp}/t= 0.5$. 
In the level crossing scenario, single-particle spectral weight is substantially redistributed when crossing from metallic to insulating solutions. 
This is seen as a jump in the single-particle gap as well as other observables indicating a first-order transition.

\begin{figure}[t!]
\centering
 \includegraphics[width=0.9\columnwidth]{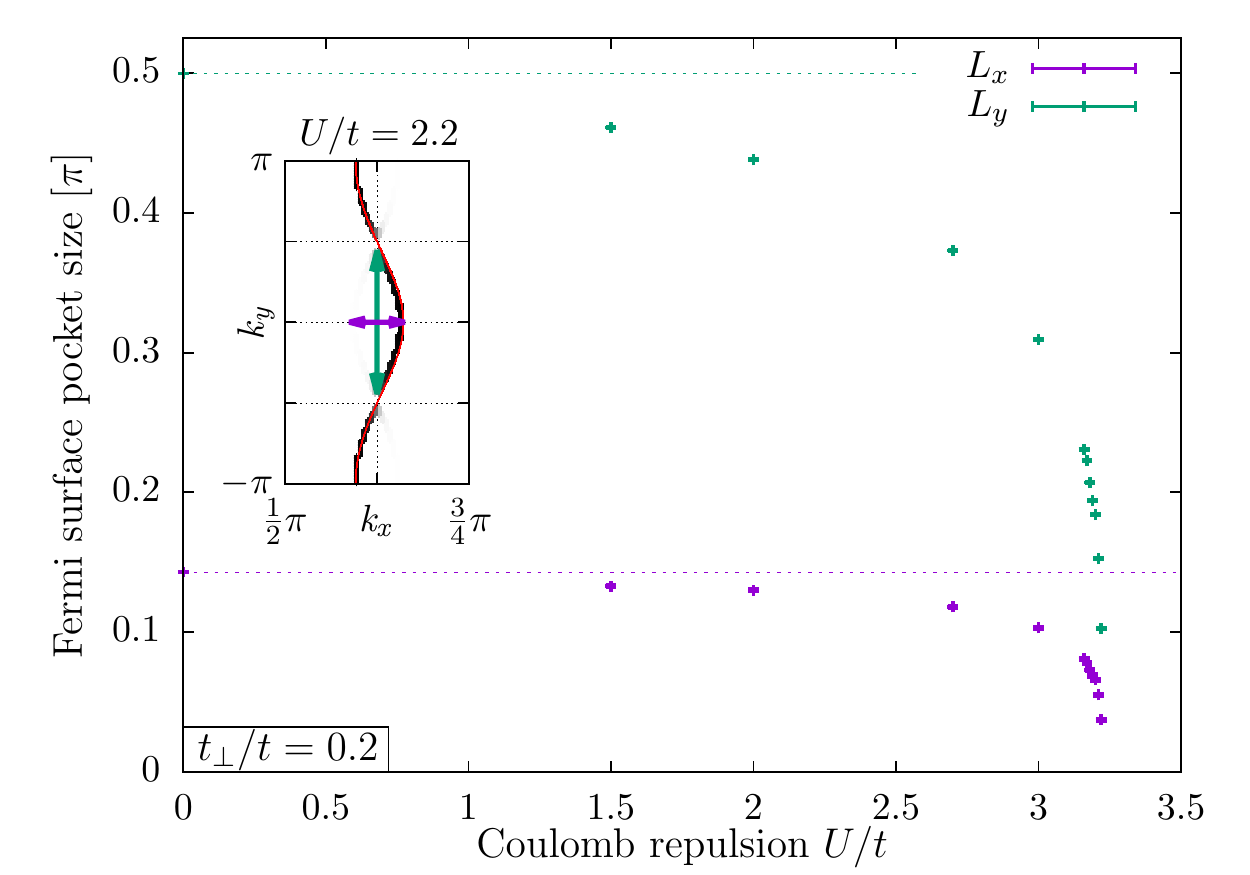} 
 \caption{(Color online)  Size of the electron pocket for $t_{\perp}/t=0.2$ as a function of $U/t$. Length of the pocket $L_y$ has been divided by a factor of two. 
 The tight-binding limit $U\rightarrow0$ is indicated by the dashed lines.}
 \label{fig: FSP_size_2}
\end{figure}

\begin{figure}[b!]
\centering
 \includegraphics[width=0.9\columnwidth]{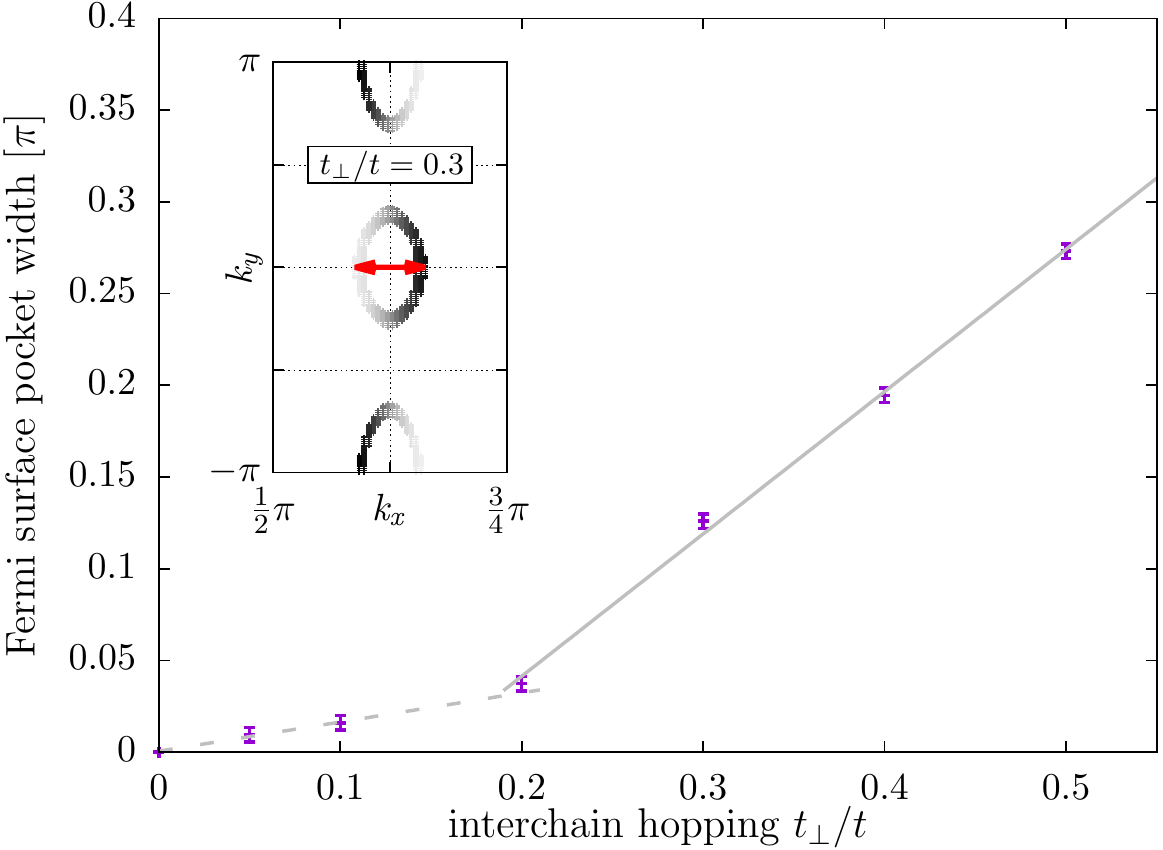} 
 \caption{(Color online) Width of the electron pocket as a function of  $t_{\perp}/t$ close to $U_c$. The location of the kink agrees with the change 
in the nature of the MIT.}
 \label{fig: FSP_size}
\end{figure}

To quantify the interaction-driven renormalization of the FS warping for $t_{\perp}/t=0.2$, we plot in Fig. \ref{fig: FSP_size_2} 
the width of the electron pocket as a function of $U/t$. For $U\ll U_c$, the main part of the FS carrying most of 
the quasiparticle weight closely follows the tight-binding dispersion. 
Increasing the interaction strength results in a measurable deviation from the tight-binding dispersion and the continuous metal to insulator 
quantum phase transition corresponds to the shrinkage of the volume of Fermi pockets on approaching the critical point $U_c$.
Since the inverse width of the hole or electron pockets defines a characteristic length scale, $\xi$, one should be able to extract the correlation 
length exponent $\nu$~\cite{IFT98S}. Thus, the analysis of the critical behavior of the volume of Fermi pockets provides an opportunity 
to address the universality class of the MIT along the quantum critical line below $t_{\perp}/t=0.2$  which is left for future.

The change in the nature (second- vs first-order) of the MIT with increasing interchain coupling is also seen in the  $t_{\perp}$ dependence 
of the pocket width as a kink at $t_{\perp}/t=0.2$, see Fig.~\ref{fig: FSP_size}. It marks the crossover between the regime where 
the volume of hole and electron Fermi pockets vanishes continuously at the second-order Mott transition and the region where the volume reduction 
of the pockets is cut off by a first-order transition.

%

\end{document}